\documentclass[aps,prl,reprint,superscriptaddress,amsmath,amssymb,
 aps,floatfix]{revtex4-1}

\usepackage{graphicx}% Include figure files
\usepackage{dcolumn}% Align table columns on decimal point
\usepackage{bm}% bold math

\usepackage{xcolor}

\usepackage[normalem]{ulem}

\newcommand{\beginsupplement}{%
       \setcounter{table}{0}
       \renewcommand{\thetable}{S\arabic{table}}%
       \setcounter{figure}{0}
       \renewcommand{\thefigure}{S\arabic{figure}}%
    }

\usepackage{xr}% one can cite things from external documnet sthen
\begin{document}

\preprint{APS/123-QED}

\title{Nonlinear analysis of the fluid-solid transition in a model for ordered biological tissues}
\author{Preeti Sahu}
\thanks{These two authors contributed equally}
\affiliation{Department of Physics, Syracuse University, Syracuse, New York 13244, USA}
 %\email{prsahu@syr.edu}
\author{Janice Kang}
\thanks{These two authors contributed equally}
 %\email{jkang@hamilton.edu }
\affiliation{Hamilton College, Clinton, New York 13323, USA}%
\author{Gonca Erdemci-Tandogan}%https://www.overleaf.com/7881599665bxkzpgkdtctj
 %\email{gerdemci@syr.edu }
 \affiliation{Department of Physics, Syracuse University, Syracuse, New York 13244, USA}
 \author{M. Lisa Manning }
 \email{mmanning@syr.edu }
\affiliation{Department of Physics, Syracuse University, Syracuse, New York 13244, USA}

\date{\today}% It is always \today, today,

\begin{abstract}
The rheology of biological tissues is important for their function, and we would like to better understand how single cells control global tissue properties such as tissue fluidity. A confluent tissue can fluidize when cells diffuse by executing a series of cell rearrangements, or T1 transitions. In a disordered 2D vertex model, the tissue fluidizes when the T1 energy barriers disappear as the target shape index approaches a critical value ($s_0^*\sim{3.81}$), and the shear modulus describing the linear response also vanishes at this same critical point. However, the ordered ground states of 2D vertex models become linearly unstable at a lower value of the target shape index (3.72)~\cite{Farhadifar2007a,Staple2010}. We investigate whether the ground states of the 2D vertex model are fluid-like or solid-like between 3.72 and 3.81 -- does the “equation of state” for these systems have two branches, like glassy particulate matter, or only one? Using four-cell and many-cell numerical simulations, we demonstrate that for a hexagonal ground state, T1 energy barriers only vanish at $\sim{3.81}$, indicating that ordered systems have the same critical point as disordered systems. We also develop a simple geometric argument that correctly predicts how non-linear stabilization disappears at $s_0^*$ in ordered systems.
\end{abstract}

\pacs{Valid PACS appear here}% PACS, the Physics and Astronomy
\maketitle

%\tableofcontents

\section{\label{sec:intro}INTRODUCTION}

% broad narrow down
%\lisa{change $p_0$ to $s_0$ everywhere, please.}

The rheological properties of a biological tissue -- how the tissue responds to stresses and strains -- and the regulation of such properties are crucial for many biological processes. 
For example, mature skin tissue typically behaves like an elastic membrane, although in processes like wound healing individual cells can change neighbors and migrate over long distances\cite{Zhang2017,Serra-Picamal2012}. Such solid-fluid transitions have recently been shown to play an important role in development~\cite{Mongera2018} and disease~\cite{Park2015b}. Thus, we would like to understand how the emergent material properties, or rheology, of the tissue is regulated by single-cell properties such as adhesivity and traction forces.

  In confluent epithelial layers where there are no gaps or overlaps between cells, it is thought that many single-cell properties can be encoded in the cell shape. For example, cells with more cadherin-based adhesion tend to share longer joint interfaces, while those with higher cortical tension tend to have smaller shared interfaces~\cite{Yamada2005, Maitre2012a}. In addition, traction forces generated by cells adhering to a substrate can also influence cell shapes. For this reason, past work has focused on a dimensionless cell shape index $s_0$ that is defined for an epithelial layer as the ratio between the cell's cross-sectional perimeter and the square root of the cell's cross-sectional area. 
  
  Extending this idea, so-called vertex models for epithelial tissue represent tissues as a network of edges and vertices, and associate a mechanical energy with the shape of each individual cell in a tessellation. Such simple models have been surprisingly successful at describing the statistics and behavior of many biological tissues~\cite{Nagai2001a, Farhadifar2007a, Chiou2012, Fletcher2017, Merkel2017}.

%while blood (a connective tissue) requires to flow like a fluid does. For processes like wound healing though, it needs to coagulate (fluidity is regulated) to make healing faster. As bulk rheology is so crucial we need to understand how single cell properties affect and regulate it. Tissues that are confluent arrangement of several cells might display rheological properties that can be directly understood/extrapolated from single cell properties like- cell shape, diffusivity, adhesivity etc.

%Rheological properties and its regulation are crucial for several biological processes. Functionality of different tissue types often hinges on it's bulk rheology for example- skin (composed of epithelial tissue) behaves like a stretchy elastic membrane while blood (a connective tissue) requires to flow like a fluid does. For processes like wound healing though, it needs to coagulate (fluidity is regulated) to make healing faster. As bulk rheology is so crucial we need to understand how single cell properties affect bulk rheology. Tissues that are confluent arrangement of several cells might display rheological properties that can be directly understood/extrapolated from single cell properties like- cell shape, diffusivity, adhesivity etc.

%introducing T1s and their importance
A further step is to understand how the single-cell shapes influence the cell dynamics and in turn the large-scale rheological properties. In tissues that exhibit a fluid-like rheology, such as those in early stages of development~\cite{Manning2010b} and in some cell cultures~\cite{Nnetu2012, Park2015b}, individual cells are able to diffuse across the tissue. In a confluent tissue with no cellular proliferation or death, the only way for a cell to diffuse is to make a series of topological rearrangements, which are called T1 transitions in 2D. During this process, an edge between two cell shrinks to zero length and then a new edge grows between two new cells, as illustrated in Fig.~\ref{fig:fullT1}(a-c). This junction remodeling results in cells changing neighbors, and many such exchanges lead to cell diffusion.

%In this rearrangement, the cell squeezes in through a neighboring pair of cells to become neighbors with a cell from the subsequent layer. In geometrical terms, the common edge shared by the neighboring cells shrinks to zero and grows in the other direction (junction remodelling). Hence T1 rearrangements are indispensable for fluidization of such a tissue.%perhaps some the geometric detailing can be moved to inland text

%nice! so why do I care?- because they are not free and the cost can tell us about the solid-fluid transition point!!
This elementary step generally requires a cell to overcome a nonlinear mechanical energy barrier. Hence, we expect the tissue to become fluid-like when the barrier associated with a T1 transition approaches zero, as this would mean that cells can diffuse and change neighbors at no energy cost. Work by Bi \textit{et al}.~\cite{Bi2015c} on homogeneous \textit{disordered} tessellations of cells has  demonstrated that the T1 energy barriers' height depends sensitively on the target shape index ($s_0$) of cells. For the 2D vertex model, this energy barrier vanishes if cells have $s_0>3.81$. Similarly, a careful analysis of the shear modulus indicates  that the tissue becomes floppy at the same critical value.

While much previous work has focused on disordered tissues, there are several examples where ordered tissues are important in biology. For example, the pattern of cells in a developing fruit fly wing becomes highly ordered at a certain stage of development; it is thought that this ordering helps to correctly orient bristles on the fly wing that are important for guiding air flow during flight~\cite{Classen2005a}. Along the same lines, sensory hairs in cochlea, which mechanically amplify low-level sound for the eventual conversion into a neural signal, develop on a highly ordered mosaic of auditory epithelial cells~\cite{McKenzie2004}. Extremely ordered structures are also found in the arrangement of lens fibre cells in vertebrate eye. It has been suggested that these ordered structures minimize light scattering by cell membrane and thereby increase transparency~\cite{Tardieu2002}. A disruption to this ordered arrangement can lead to cataracts~\cite{Cooper2008}. %\lisa{Preeti, please add some citations and read those citations to see if there are other examples.}  
In addition, scientists are currently developing biomimetic cellular systems~\cite{Pontani2016}, where ordered systems could be engineered. Ordered tessellations have also been studied extensively from a theoretical point of view, and help to provide a deeper understanding of the role disorder plays in altering the rheological properties of a tissue. 

Within the framework of vertex models, Farhadifar \textit{et al}.~\cite{Farhadifar2007a} and then Staple \textit{et al}.~\cite{Staple2010} performed a beautiful and comprehensive investigation of ordered tessellations. Using linear stability arguments, they demonstrated  that the shear modulus of ordered ground states of the 2D vertex model disappear for all $s_0>3.722 = s_{hex}$, the perimeter to area ratio of a regular hexagon. In other words, the energy landscape is locally flat for $s_0>3.722$. In normal materials this typically suggests that the nonlinear energy barriers corresponding to structural rearrangements also cost no energy. However the energy landscape of vertex models is known to have cusps~\cite{Sussman2018} and so it is possible that energy barriers still exist even though the linear response is flat.

%\preeti{In other words, hexagonal tissues, in this shape regime, reside on a locally flat energy landscape since they are at a global minima. It can however jump to a neighboring local minimum by performing a cellular rearrangement. This would generically mean- climbing up on non-linear energy barriers. When these energy barriers go to zero, cellular rearrangements can occur seamlessly and the bulk tissue can behave like a fluid. Hence, in order to characterize a solid-fluid transition in ordered systems, one needs to explore the non-linear regime in addition to linear response.}

This presents an interesting open question, which is -- what is the nature of the \emph{nonlinear} mechanical response of ordered tissues. One possibility is that the energy barriers also vanish for $s_0>3.722$. This would suggest that there are two branches to the equation of state for cellular materials: an ordered branch that becomes floppy at $s_0 =3.722$, and a disordered branch that becomes floppy at  $s_0 =3.81$. This would be similar to the scenario in jammed particulate matter, where there is also an ordered and disordered branch and the control parameter is the packing fraction, instead of cell shape ~\cite{Kamien2007}. An alternate possibility is that the energy barriers for ordered tessellations vanish at some other value of $s_0$, possibly even at the same value as in the disordered solid. This would also mean that there is a strong discrepancy between the linear and nonlinear response in the non-ordered vertex model, hinting at the possibility of non-analyticity.

%At the same critical value, the tissue becomes floppy and has a fluid-like response to linear shear ie the shear modulus vanishes. On the other hand, the scenario for \textit{ordered} tessellation of cells is not so well understood. While T1 barriers are not computed for such tissues \cite{Farhadifar2007a,Staple2010}, have found that the linear response of ordered ground states of 2D vertex model, becomes fluid-like for $p_0>3.722$. We want to understand if this behaviour persists for non-linear regime (which is the case for T1 transitions). If it does, then we could establish a similarity with particulate glasses, since the transition point is different for disordered vrs ordered tessellation of cells. From modeling perspective, knowing in both linear and non-linear regime gives a more complete picture of fluid-solid transition in ordered tissues.

% Lisa's part

%what is coming next 
In this work we first study the energy barriers for a bulk ordered tissue. We find that although tissues with $s_0>3.722$ are linearly unstable, they are non-linearly stabilized up to $s_0^*=3.81$, which establishes that both ordered and disordered tissues have the same transition point. We study a simple one-cell geometric construction that describes this process and quantitatively predicts features of nonlinear stabilization.

\section{\label{sec:method}Model and Methods}
To find the transition point based on T1 energy barriers, we simulate a 2D confluent monolayer using a Vertex model~\cite{Nagai2001a, Farhadifar2007a,Teleman2007,Staple2010,Manning2010b,Hilgenfeldt2008, Chiou2012,Bi2015c,Fletcher2014}. 
%\lisa{Preeti, please add lots of cites here}.
%A successful class of models to describe confluent tissues on the cellular scale are the Vertex models where 
Vertex models describe the energy of a 2D tissue containing N cells as
\begin{equation}\label{energy}
    E_j= \sum_j^N K_{Aj} (A_j - A_{0j})^2+K_{Pj} (P_j -P_{0j})^2.
\end{equation}
Here the first term represents cell volume incompressibility, and $A_j$ and $A_{0j}$ are the actual and preferred areas of cell $j$. The second term models actomyosin contractility and adhesion between the cells, where $P_j$ and $P_{0j}$ are the actual and preferred perimeter of cell $j$. $K_{Aj}$ and $K_{Pj}$ are the area and perimeter moduli, respectively. We consider the homogeneous case where all single-cell properties are equal ($K_{Aj}=K_A, K_{Pj}=K_P,A_{0j}=A_0,P_{0j}=P_0$). The energy functional in Eq.~\ref{energy} can be non-dimensionalized in length $\sqrt{A_0}$ resulting an effective target shape index $s_0=P_0/\sqrt{A_0}$ which has been shown to control rigidity or glass-like transitions in such systems \cite{Bi2015c}.

% each $j^{th}$ cell has an energy that depends only on its area($A_j$) and perimeter($P_j$) as follows- 
% $$E_j= K_A (A_j - A_{0})^2+K_P (P_j -P_{0})^2$$
% Here $A_{0}$ and $P_{0}$ are the preferred- area and perimeter respectively. $A_{0}$ is set to unity in this project making the natural length-scale $\sqrt{A_{0}}=1$ as well. Deviation from these preferred values costs a quadratic penalty that is tuned by the area and perimeter stiffness- $K_A$ and $K_P$. $K_A$ is set to 100 to mimic incompressibility. Total energy of the tissue is then defined as $E=\sum E_{j}$. The dimensionless quantity- target shape index $p_0\equiv P_0/\sqrt{A_0}$ has been shown to control rigidity or glass-like transitions in such systems. To compute the T1 energy barriers we need to minimize the energy of this tessellation. Each vertex of the polygonal cells is a degree of freedom (dof) in the model. These dofs rearrange according to the minimization procedure to reach an energy minimum. %One needs to take care of cellular rearrangements via T1 transitions explicitly. If an edgelength falls below some threshold length $l_c$, then we rotate the edge by $\pi/2$ and reorient the surrounding cells to undergo a local neighbor-exchange operation between the four cells involved. Unless otherwise specified $l_c$ is set to 0.04. 
              %put the equations and talk in details about vertex model and proabbaly FIRE minimization technique a bit?
              
\begin{figure}[!htbp]
\centering
\includegraphics[width=0.9\columnwidth]{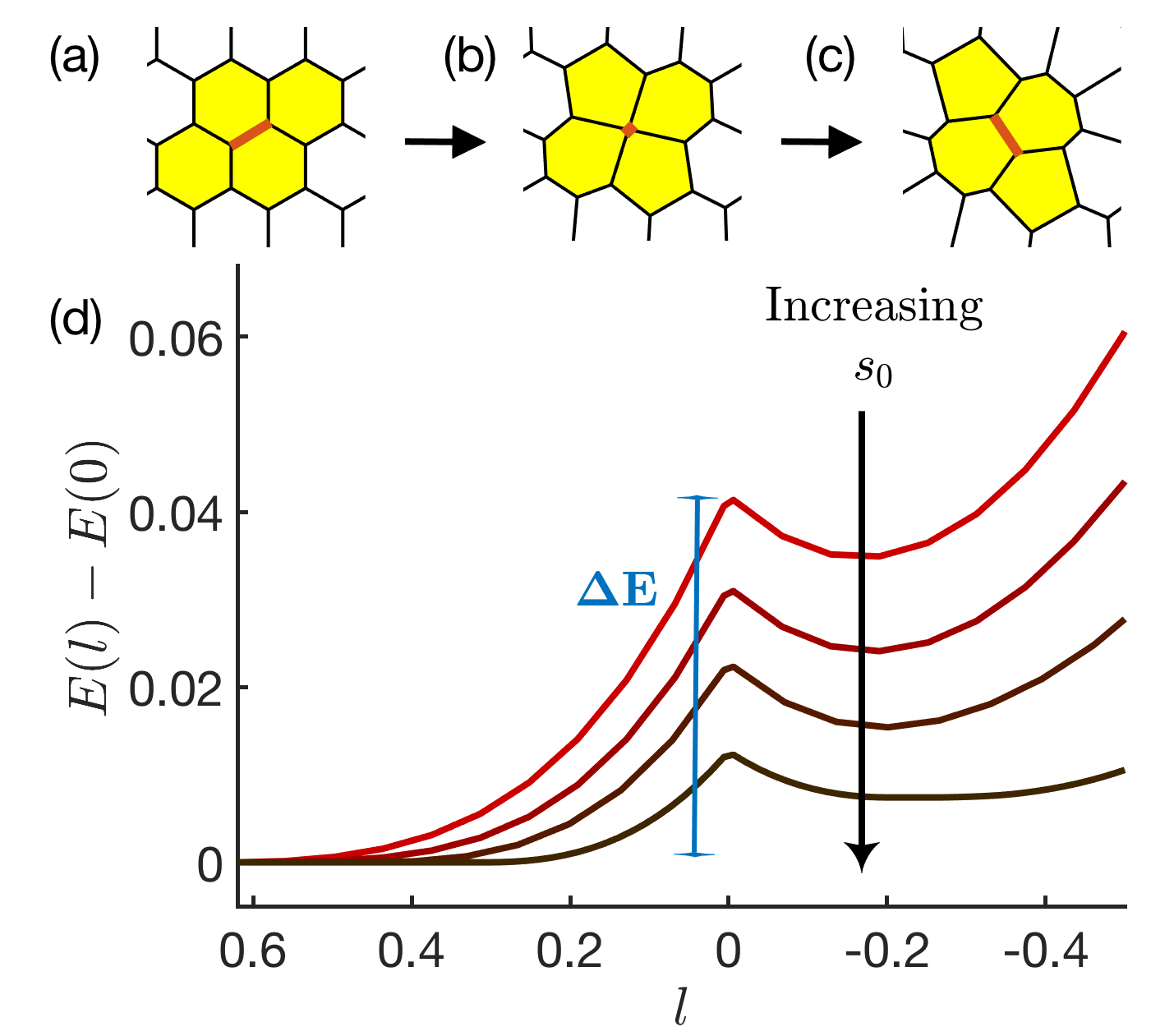}
  \caption{\textbf{Energetics of an ordered T1 transition}:(a) A T1 edge, highlighted in red, at its rest length, (b) the T1 edge shrinks to zero length (c) the T1 edge rotates by $90^{\circ}$ and is then expanded. (d) A typical energy profile across the T1 transition plotted with respect to the T1 edge length $l$ during of T1 junction remodelling, for $s_0$ 3.71, 3.72, 3.73 and 3.75 (light red to dark red). The T1 energy barrier for the lowest $s_0$ is the peak height (highlighted in blue vertical line). }
  \label{fig:fullT1}
\end{figure}

Cell neighbor exchanges happen through T1 transitions. A typical T1 process is shown in Fig.~\ref{fig:fullT1}(a-c). As the T1 edge $l$ shrinks from its rest length, $l_0$, (Fig.~\ref{fig:fullT1}(a)), it eventually achieves a \textit{transition state} at $l=0$ with a 4-fold vertex where all 4 cells are neighbors (Fig.~\ref{fig:fullT1} (b)). This is followed by a $90^{\circ}$ reorientation of the T1 edge and expansion along the new direction (Fig.~\ref{fig:fullT1} (c)). We find that the mechanical energy of the tissue is maximized at the transition state with the 4-fold coordinated vertex. As in previous work, we describe the difference between the initial energy and maximum energy as an energy barrier that must be overcome for cells to change neighbors. In analogy with activation energies required for diffusion in Arrhenius processes, we can then think of the T1 edge-length ($l$) as a reaction coordinate \cite{Bi2014,Bi2015c}. 

 We focus on the first part of the T1 process for the rest of this paper, as this is sufficient to compute the energy barrier (shown in blue vertical line in Fig.~\ref{fig:fullT1}). We choose the sign convention as positive for this part of the transition, which is different from the convention used for $l$ in work that studies both sides of the transition~\cite{Bi2015c}.

The difference between the peak energy $E_f$ and the initial energy $E_i$ gives the T1 energy barrier (Fig.~\ref{fig:fullT1} vertical line in blue),

\begin{equation}
\label{eq:barrier}
   \Delta E (l)=E_f-E_i=E(l)-E(0). 
\end{equation}

%{\bf Preeti, you use $l_c$ and $dt$ in figure captions but they are not defined in the text. I explain the $l_c$ in the following. Could you add an explanation for $dt$ along with the types of dynamics/equation of motion you use. Similary, could you explain the choice of $K_A$ value? (Why 100?)}
For the bulk simulations, we use the open source cellGPU code~\cite{Sussman2017}. A FIRE minimization protocol~\cite{Bitzek2006} is used for bulk energy minimization. The initial FIRE step, $dt$, is set to 0.01. The T1 protocol is such that a T1 transition forms whenever the distance between two vertices is less than a critical value, {$l_c$}. We chose $l_c=0.006$ for the ordered tissue simulations.

%what are we doing differently for disordered system
As discussed in the ESI, we apply the same procedure to compute the transition point in disordered systems. Unlike ordered systems, which have a unique hexagonal initialization, in a disordered systems we average the energy barrier profile over different initializations. See the ESI for more details. 
%{\bf(perhaps do not refer the figure as it jumps from Fig.1 to Fig.7 in the text.)}
%using a higher T1 cut-off of $l_c=0.15$ to reduce T1s happening elsewhere after equilibration.

%\preeti{From recent work (my work in 2d mixtures studied Deff for fixed temperature as function of $K_A$ in disordered system), 

Recent work ~\cite{Sahu2019} has shown that the transition point in vertex models is unaffected by the choice of $K_A$. Here, we choose $K_A=100$, which enforces that cells remain close to their preferred area $A_0=1$.

\subsection{\label{sec:manycell}Many-cell system}

To test the transition point of ordered tissues subject to a specific non-linear perturbation, we construct a rectangular periodic box that can accommodate an integer number of hexagons, with a length-to-width ratio of $\frac{3m}{2\sqrt{3}n}$,
%along the horizontal and vertical axis using Vertex model. The box is also ensured to be large enough that a T1 does not interact with its own image. The ratio of the length and width of such a box is found to be , 
where $m$ is the number of hexagons along the vertical axis and $n$ is the number of hexagons along the horizontal axis. We investigate small systems with $N=90$ such that $n=9$ and $m=10$, simulated using cellGPU code. 

%To execute a T1 process, we randomly chose an edge in this initial configuration of the hexagonal lattice.

%One can then feed in a hexagonal lattice as the initial configuration and plant a T1 process in a randomly chosen edge after equilibration. %The threshold length $l_c$ is reduced to a very small value (0.005) to prevent T1s from happening elsewhere in the system. 

\begin{figure}[!htbp]
\centering
\includegraphics[width=0.9\columnwidth]{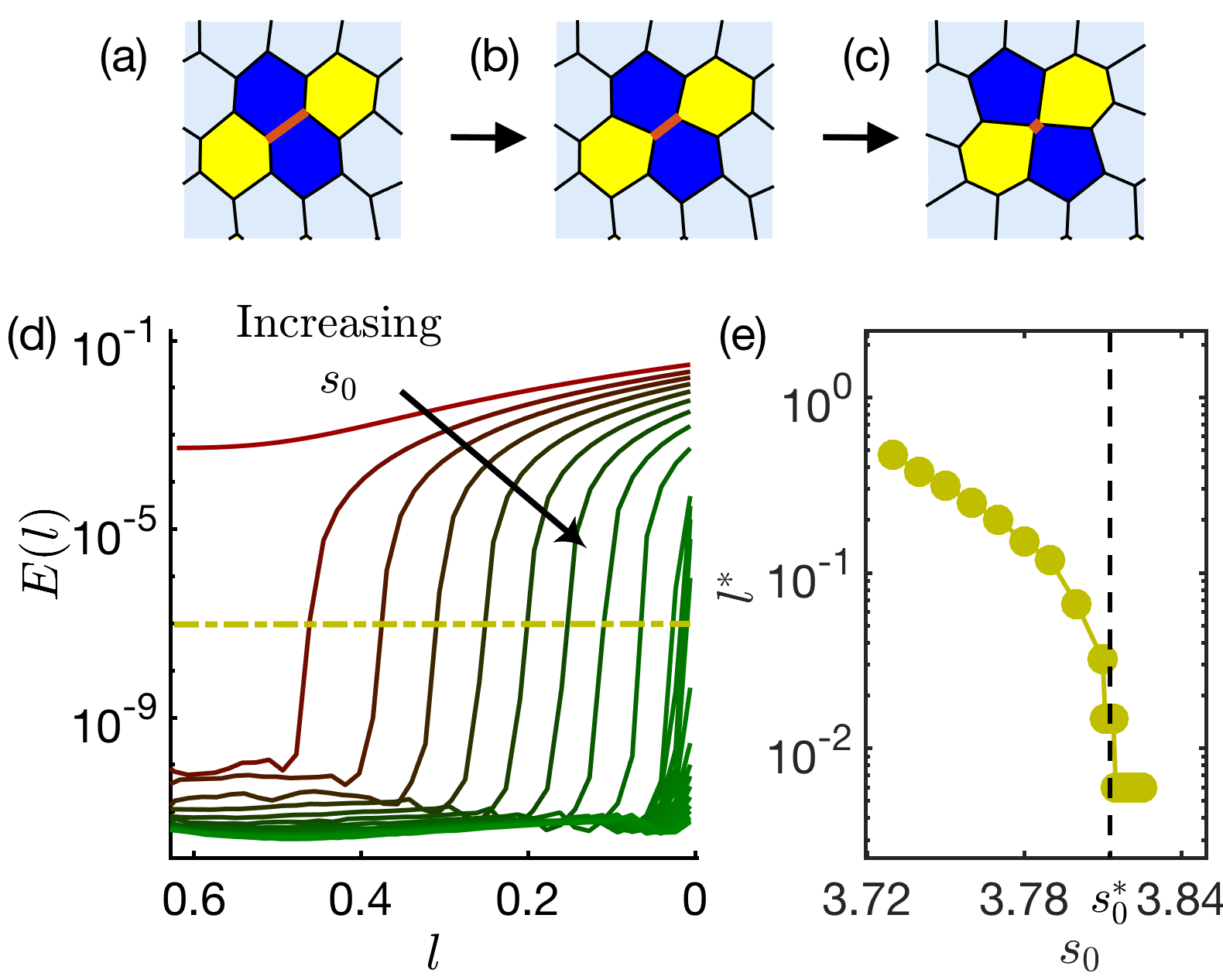}
  \caption{\textbf{ Many-cell energy profile}: (a-c) A snapshot from an ordered tessellation of 90 cells with $s_0=3.76$. A randomly chosen edge (highlighted in red), shrinks to zero length (left to right as directed by the arrows). (d) In this process, the total energy of the tissue, $E$, is plotted against the shrinking T1 edgelength $l$ for increasing values of $s_0$ (3.72 to 3.81 in steps of 0.01 and 3.810 to 3.825 in steps of 0.001) varying from red to green. The energy cut-off is shown by yellow dash-dot line. (e) The critical edgelength $l^*$ associated to the cut-off shown in (d) is plotted for each $s_0$ value in yellow circles. The dashed line indicates critical $s_0^*$ found for disordered tissues. 
  %Simulations have other parameters- the T1 threshold length $l_c=0.006$, integration time step $dt=0.01$ and $K_A=100$. 
  %{\bf (Why did we pick 100? Should the results be similar for $K_A=1$ as this is the Vertex model?)}
  }
  \label{fig:manycell}
\end{figure}

A random edge of the ordered confluent tissue is chosen to undergo a T1 transition, and the energy profile is analyzed across different $s_0$ values. A typical T1 edge, with its neighbourhood, is shown in Fig.~\ref{fig:manycell} (a) along with energy profiles for different $s_0$ values (Fig.~\ref{fig:manycell}(d)). For values of $s_0 < 3.722$, any perturbations of edge lengths costs finite energy, as illustrated by the red curve in Fig.~\ref{fig:manycell}(d). For values of $s_0 > 3.722$, we find that small perturbations of $l$ require zero energy 
%(far right of Fig~\ref{} (d)) {\bf (new plot?)} \preeti{I am sorry I did not understand this comment Gonca}, 
as previously predicted ~\cite{Farhadifar2007a, Staple2010} using linear response. This is indicated by values of $E(l)$ near zero on the left-hand side of Fig.~\ref{fig:manycell}(d). %As one might expect from Farhadifar et al.'s prediction\cite{Farhadifar2007a}, the linear regime in the far right of the plot that depicts infinitesimal perturbation from the rest-length configuration, costs zero energy (below the chosen cutoff of $10^{-7}$) as soon as $p_0$ value drops from 3.72 to 3.73.
But as the T1 proceeds further, the energy becomes finite at a critical lengthscale $l^*$. In practice, we identify $l^*$ as the point at which the energy first rises above a cutoff value of $10^{-7}$ shown by the dashed yellow line in Fig.~\ref{fig:manycell}(d).
%
%the system gets non-linearly stabilized i.e. it does not remain below the cut-off all throughout. In fact, there is a new length-scale $l^*$, shrinking below which costs energy. We compute this $l^*$ for each $p_0$ by computing the midpoint of lengths before and after energy shoots above the cutoff. One might expect from the energy profile that the $l^*$ diminishes as $p_0$ approaches 3.813. By plotting $l^*$ as a function of $p_0$ (Fig. \ref{fig:manycell}(b)), we confirm that
We find that $l^*$ diminishes with increasing $s_0$, and drops to zero at $s_0^* = 3.81$, which is the same value identified in disordered systems, as shown in Fig.~\ref{fig:manycell}(e). We note that the lowest value of $l^*$ accessible in our simulations is limited by the T1 threshold length, $l_c=0.006$.

%For $p_0<3.722$, as expected the energy is non-zero and increases with shrinking T1 edge. Only after $p_0$ of 3.722 can one expect the energy to start vanishing as they become linearly unstable. We verify that an infinitesimal change from the rest length configuration costs zero energy for $p_0>3.722$. The inset in \ref{fig:manycell} confirms this dramatic fall of energy as $p_0$ is increased from 3.72 to 3.73. But as the T1 proceeds further into non-linear regime, the energy does not remain at zero. In fact, depending upon the $p_0$, it becomes non-zero after a threshold value of $l$*. This $l$* goes to zero as $p_0$ approaches $\sim 3.813$ as shown later in \ref{fig:overlay}. Therefore for bulk system, we show that the energy barriers remain non-zero up till $p_0\sim 3.813$.

%But can we understand this better by looking at a mean-field candidate for this process?

\subsection{\label{sec:4cell}Four-cell system}
\begin{figure}[!htbp]
\centering
\includegraphics[width=0.9\columnwidth]{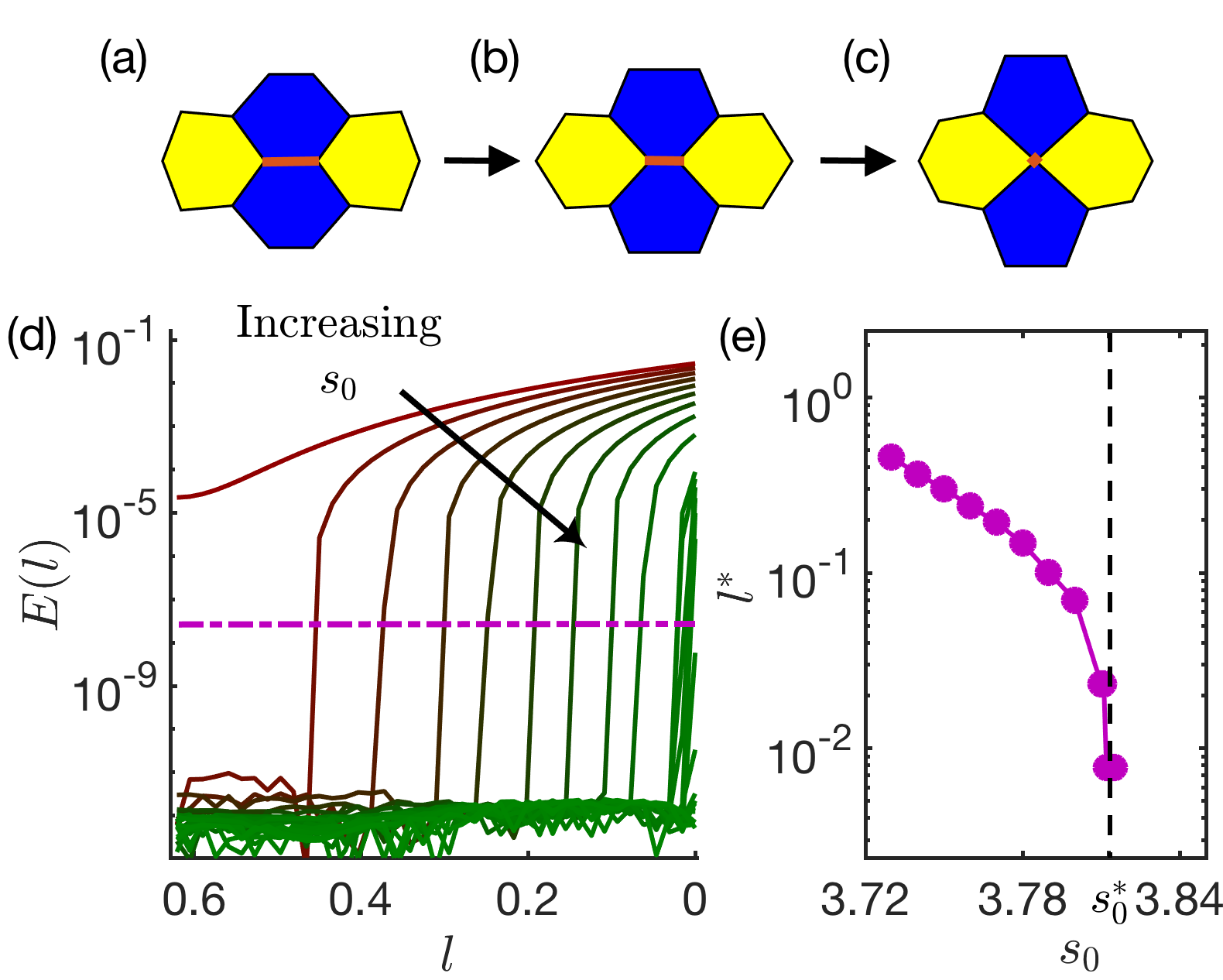}
  \caption{\textbf{ Four-cell energy profile}: (a-c) In an ordered initial configuration of 4 cells with, the T1 edge, shrinks to zero length (left to right as directed by the arrows). (d) In this process, the total energy of the 4-cell unit, $E$, is plotted against the shrinking T1 edgelength $l$ for increasing values of $s_0$ (3.72 to 3.81 in steps of 0.01 and 3.810 to 3.825 in steps of 0.001) varying from red to green. The cut-off for the energy is shown by the magenta dash-dot line. (e) The critical edgelength $l^*$ associated to the cut-off shown in (d) is plotted for each $s_0$ value in the magenta circles. The dashed line indicates critical $s_0^*$ found for disordered tissues.}
  \label{fig:4cell}
\end{figure}

%Although for bulk ordered tissue, we show that the transition point is the same as that of disordered systems, we want to understand the reason behind this non-linear stabilization. 
To better understand the origin of $l^*$, we next study a simple mean-field model for a T1 process over a 4-cell system composed of hexagons. For such a system, we minimize Eq.~\ref{energy} for every value of $l$ between the edge rest length and zero across different values of $s_0$ as shown in Fig. \ref{fig:4cell}. To perform such minimizations, we must constrain the geometry using symmetry considerations. Specifically, a 4-cell system comprised of hexagons is expected to have a total of 16 vertices and hence 32 degrees of freedom (DOFs). If we assume symmetry about the x- and y-axes, this reduces the system to eight orthogonal DOFs.  For each energy minimization step the T1 length is fixed, resulting in seven DOFs. Since there are 8 constraints (2 on each cell) imposed by the energy functional, we can solve the resulting system of equations uniquely.

A typical T1 edge is shown in Fig.~\ref{fig:4cell}(a-c) along with energy profiles for different $s_0$ values shown in Fig.~\ref{fig:4cell}(d). Similar to the many-cell system, infinitesimal perturbations cost energy for $s_0 < 3.722$. For $s_0 > 3.722$, perturbing the system a small amount costs zero energy, but as the T1 proceeds further into non-linear regime, the energy becomes non-zero after a threshold value of $l^*$. This $l^*$ goes to zero as $s_0$ approaches $\sim 3.813$ as shown in Fig.~\ref{fig:4cell}(e). We observe that the energy profile is qualitatively similar to that of a many-cell system (Fig.~\ref{fig:manycell}) which confirms that a simple 4-cell unit is a suitable mean-field model for T1 processes in ordered tissues. 

%Symmetry is an extra feature that we put in, that gives rise to symmetric polygonal configurations which need be the case in general (as seen in many-cell T1 snapshots\ref{fig:manycell}(a)). But when it comes to the non-floppy port 

%One potential mean field representation for this process can be a 4-cell unit involved in a T1 itself.

%{\bf Edit and move the following red parts to discussion?}\textcolor{red}{
%We assume that high $K_A$ and symmetry about x and y axis mimics a mean-field-like interaction with the neighbours.  With the added constraint of symmetry about x and y axis it reduces down to 8 independent dofs. The T1 edge shrinkage is done by fixing the T1-edge-dofs at a new smaller length ($l$) and minimizing the system to acquire a final state that brings the system to a minima. This final state then acts as the initial configuration for the subsequent minimization. This procedure helps prevent unphysical folding up of polygons as well. Since this step requires pinning another dof, the total number of independent dofs reduce to 7. Thus for every minimization, the number of independent dofs is outnumbered by area and perimeter constraints (2 per cell). Allowing for asymmetric configurations makes the system very floppy. }

%This system is simulated using MATLAB and fmincon (Sequential Quadratic Programming- sqp) performs the minimization for every value of l starting from its rest length to 0 across different values of $p_0$ as shown in Fig:\ref{fig:4cell}. 

%For the 4-cell system, we shrank the T1 edge, l, for each Po until the two initially non-adjacent cells became neighbors. This shrinkage process was repeated recursively and the energy of the cells based on Eq. (1) was minimized in the course of the T1 transition. 

%A typical T1 edge is shown in \ref{fig:4cell}(a) along with energy profiles for different $p_0$ values. Just like, many-cell system, only after $p_0$ of 3.722 does the energy starts vanishing (as confirmed by the dramatic fall of energy as $p_0$ is increased from 3.72 to 3.73 in \ref{fig:4cell})(a). But as T1 proceeds further into non-linear regime, the energy starts becoming non-zero after a threshold value of $l$*. This $l$* goes to zero as $p_0$ approaches $\sim 3.813$ as shown in \ref{fig:4cell}(b). We observe that the energy profile is qualitatively no different from that of a many-cell system \ref{sec:manycell} which confirms that it is a suitable mean field element for T1 processes in ordered tissues. %Symmetry is an extra feature that we put in, that gives rise to symmetric polygonal configurations which need be the case in general (as seen in many-cell T1 snapshots\ref{fig:manycell}(a)). But when it comes to the non-floppy port 

\bigskip

%We will refer to the cells that share a T1 edge at the initial stage as central cells, while the remaining two outer cells are referred to as peripheral cells. To analyze which cells in this 4-cell unit contribute the most to the energy barrier, we plot the ratio of contribution from both types [TO DO]. Geometrically the central cells transform from a hexagon to a pentagon. While there are many ways to undergo this transformation while keeping the area of the polygon intact, we concentrate on the following geometric mechanism to study a single cell system.

%as this reproduces the $l^* ~ p_0$ scaling perfectly.

\subsection{\label{sec:mechanism2}Single cell prediction}

In both many-cell and 4-cell systems, the ordered polygons that undergo a T1 transition start out as perfect hexagons but become pentagons as the edgelength ($l$) shrinks to zero (Fig.~\ref{fig:manycell} (c) and Fig.~\ref{fig:4cell} (c)). For disordered systems, the formation of a pentagon was proposed as a mean-field lower bound on the T1 transition point previously by Bi \textit{et al}~\cite{Bi2015c}.

\begin{figure}[!t]
\centering
\includegraphics[width=\columnwidth]{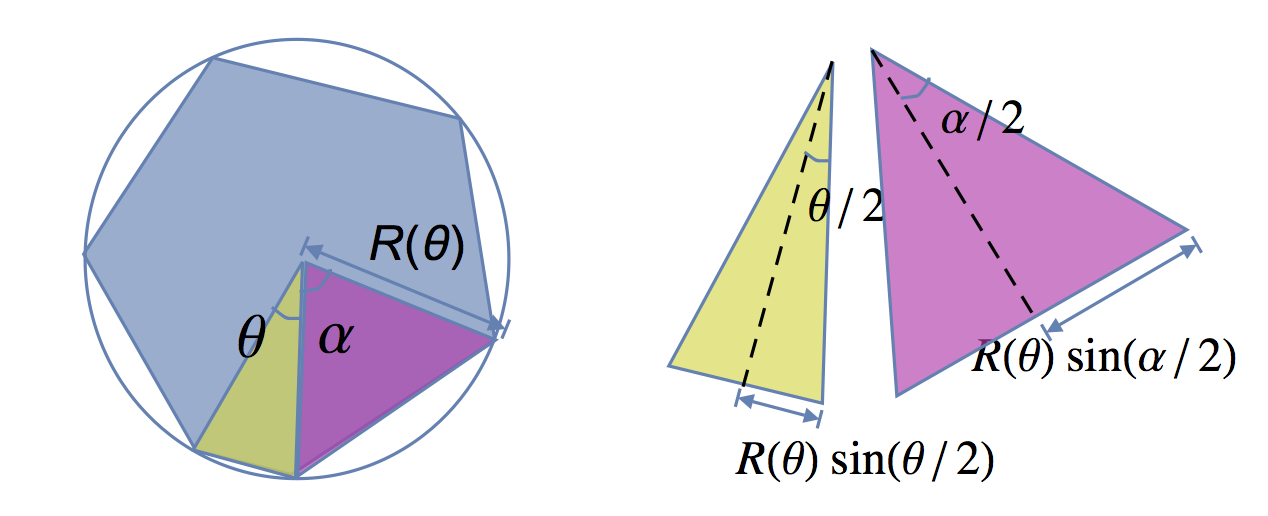}
  \caption{\textbf{A geometric mechanism for formation of a uniform pentagon}: The 6-sided polygon has five  sides equal to each other and one that is allowed to be different subjected to the constraints that the polygon lies on a circle and its area remains unity. The angles correspond to two different types of sides ($\alpha$ and $ \theta$) are highlighted in pink and green.}
  \label{fig:model3}
\end{figure}

Here we construct a geometric ansatz to predict the T1 edgelength $(l^{*})$ at which the energy barrier becomes non-zero. We restrict ourselves to study a polygon whose vertices lie on a circle of radius $R$ (Fig. \ref{fig:model3} (a)). This constraint is a simple way to enforce that the polygon remain roughly isotropic, consistent with our observations from simulations. To model the ordered case, we enforce that the polygon has six sides, one of which is constrained to shrink and subtends an angle $\theta$ at the center. We assume the remaining sides adjust themselves to be of equal length, which minimizes the remaining perimeter subject to having one constrained edge, as illustrated in Fig. \ref{fig:model3}. We can then study the perimeter change of this polygon as it transforms from a uniform hexagon to a uniform pentagon. We constrain the area of the polygon to unity to account for incompressibility of cells.

The area of the polygon can be written in terms of the area of six triangles that make up the polygon. Five of them are congruent to each other, since they subtend the same angle $\alpha$ at the center and the sides are of length $R$ (triangle $\Delta_{\alpha}$, labelled in violet in  Fig. \ref{fig:model3}). The leftover triangle subtends angle $\theta$ at the center and will be referred to as $\Delta_{\theta}$. 

The area constraint ensures $5 Ar(\Delta_{\alpha})+ Ar(\Delta_{\theta})=1$. Substituting the area in terms of angles and radius R, the radius of the circle is determined as a function of $\theta$: 

$$R(\theta)=\sqrt{\dfrac{2}{\sin(\theta)+5 \sin(\alpha))}}$$ where $ \theta+5 \alpha=2\pi$.

%Plotting the function shows that the radius increases slightly as $\theta$ decreases from $\pi/3$ (uniform hexagon).
%reference to the energy functional for why do we need to find perimeter

\begin{figure}
\centering
\includegraphics[width=0.9\columnwidth]{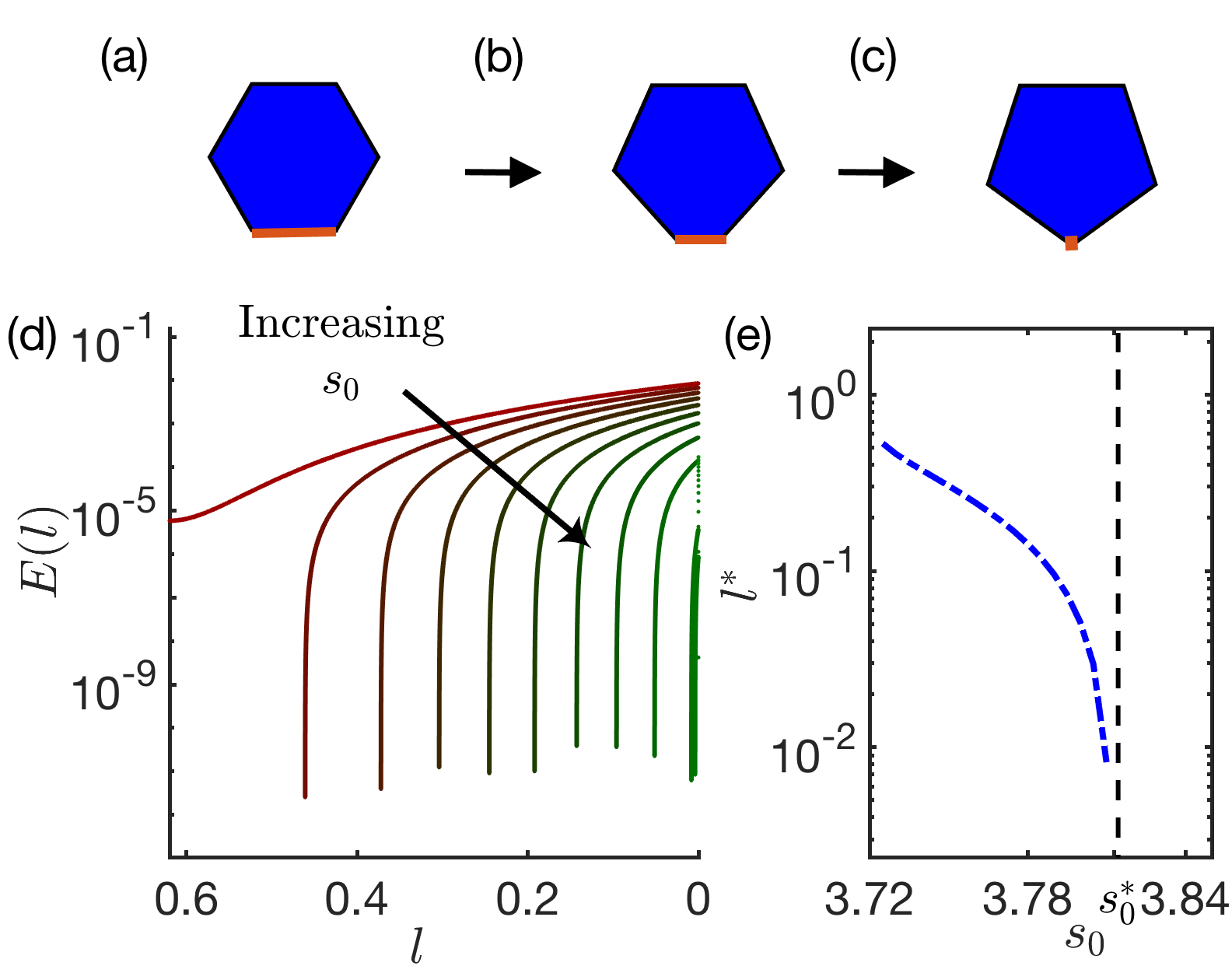}
  \caption{\textbf{ Single-cell energy profile}: (a-c) For a single cell inscribed on a circle, the T1 edge (highlighted in red) shrinks to zero length (right to left as directed by the arrows). (d) In this process, the total energy $E$ is plotted against the shrinking T1 edgelength $l$, for increasing values of $s_0$ (3.72 to 3.81 in steps of 0.01) varying from red to green. (e) The critical edgelength $l^*$ associated to the drop shown in (d) is plotted for each $s_0$ value in blue dot-dashed line. The blue dashed line indicates critical $s_0^*$ found for disordered tissues.}
  \label{fig:1cell}
\end{figure}

\begin{figure}[!htbp]
\centering
\includegraphics[width=0.75\columnwidth]{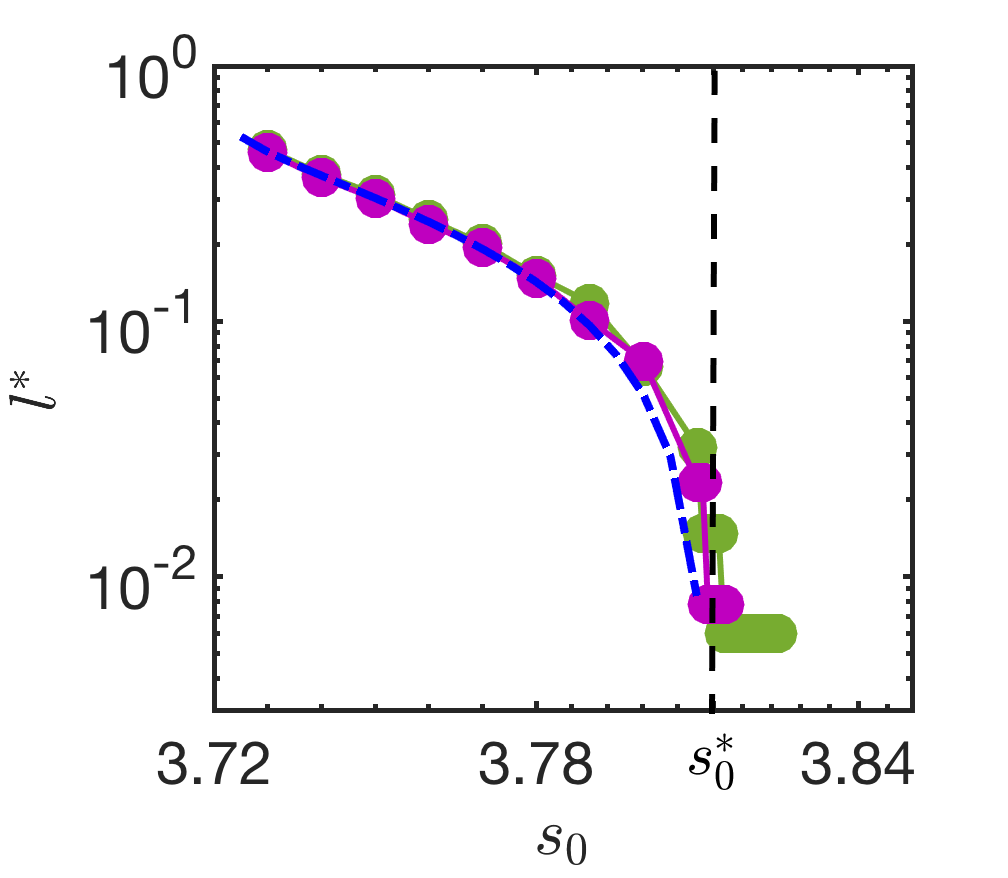}
  \caption{\textbf{Non-linear stabilization seen in ordered bulk systems can be produced in 4-cell system and single cell model}: Critical edgelength $l^*$ plotted against $s_0$ is superimposed for both- many-cell (green circles) and 4-cell systems (magenta circles). In addition, the analytical prediction from the geometric mechanism explained in the text is shown in blue dashed line. }
  \label{fig:overlay1}
\end{figure}

%Now that we have the radius as a function of the perturbation angle ($\theta$), we can compute the perimeter (P) of the polygon as-

Adding all the edgelengths, the total perimeter P, of the polygon is $P(\theta)=2 R(\theta)\{\sin(\theta/2)+5\sin(\alpha/2)\}=\sqrt{\dfrac{2}{\sin(\theta)+5 \sin(\alpha)}}2\{\sin(\theta/2)+5\sin(\alpha/2)\}$.

For this T1 process, the edge facing $\theta$ mimics the T1 edge that shrinks to zero as shown in Fig.~\ref{fig:1cell}(a). This T1 edge-length can be easily determined from $\theta$ as $l(\theta)=2R(\theta) \sin(\theta/2)$. 

For a cell of unit area the total vertex energy depends only on the deviation of the perimeter from its target value.  The target perimeter equals the actual perimeter when the angle $\theta^*$ associated with a T1 edgelength $l^{*}$ satisfies the following analytic equation:

\begin{equation}
 P(\theta^{*})=P_0 = \sqrt{\dfrac{2}{\sin(\theta^{*})+5 \sin(\alpha^{*})}}2\{\sin(\frac{\theta^{*}}{2})+5\sin(\frac{\alpha^{*}}{2})\}.
 \end{equation}

For each value of $P_0$, this equation then identifies the $l^*$ at which the energy barrier goes to zero, as shown in Fig.~\ref{fig:1cell}(d). These results are quantitatively consistent with the results for $l^*$ for the 4-cell and bulk simulations, demonstrating that a very simple geometric ansatz predicts the onset of nonlinear stabilization in the ordered vertex model. All three models exhibit very similar behavior, as shown in Fig.~\ref{fig:overlay1}, with $l^*$ dropping to zero when $l^* \sim 3.81$. 
%In addition we show the $l^*$ behaviour for disordered systems as well. The details of its computation are explained in Supp Sec\ref{sec:SI}.

\section{\label{sec:discussion}Discussion and Conclusion}

%\lisa{Still need to add edits here}
%Things to add to the discussion
%\begin{enumerate}
%    \item {Crystallization and behavior of this system could be weird under shear}
%    \item{Connection to Moshe, Marchetti, Bowick results}
%    \item{Connection to Isogonal modes work by Shraiman et al?}
%\end{enumerate}

We have demonstrated that the ordered ground states of the frequently-used 2D vertex model for biological tissues are \textbf{stable} with respect to localized cell rearrangements when the target shape parameter $s_0$ is between $3.72$ and $3.81$.  This is surprising, as previous analytic calculations for the linear response highlights that the ordered states become linearly unstable for all $s_0$ values greater than $3.72$~\cite{Farhadifar2007a,Staple2010}.

We demonstrate this nonlinear stabilization in a full simulation of the vertex model, and also in two toy models, one of which is analytically tractable. In all three models, we find that for values of $s_0$ between $3.72$ and $3.81$, small perturbations to the structure cost zero energy, in line with previous calculations of linear response. However, there is a finite scale of perturbation at which the energy suddenly becomes non-zero. In ordered systems, we characterize this behavior in terms of the edge-length $l^*$ at which the energy first becomes non-zero, and find that $l^*$ decreases monotonically from the ordered edge length $l_0$ at $s_0 = 3.72$ to zero at $s_0 \sim 3.81$.  In the simplest analytically tractable and purely geometric model, we see that $l^*$ vanishes precisely at $s_0 \sim 3.81$ because that is the point at which an isotropic pentagon costs zero energy.

As discussed in the Supplementary materials, a very similar analysis can be performed on disordered configurations of the 2D vertex model. While the data is noisier due to the disorder in edge length, it is clear that in disordered tissues the smallest values of $l^*$ remains on the order of the average edge length in the tissue for all $s_0 < 3.81$, and drops precipitously to zero for $s_0 > 3.81$. This Heavyside-function-like behavior is consistent with the hypothesis that disordered tissues also destabilize when it is possible for an isotropic pentagon to form at zero cost, as postulated previously~\cite{Bi2015c}. An interesting direction for future work would be to carefully characterize how the statistics of short edge-lengths and $l^*$s vary as a function of system size and model parameters in disordered systems, extending previous work demonstrating the importance of edge length statistics to rigidity in Vertex models~\cite{Kim2018}.

Overall, this result is interesting because it suggests that unlike particulate glassy materials, where there are two branches to the equation of state associated with ordered and disordered states~\cite{Kamien2007}, vertex models are ultimately destabilized at the same point (or at least very nearly the same point) on the state diagram, at $s_0 \sim 3.81$, regardless of the degree of disorder.

This deep connection between ordered and disordered states is only possible because the potential energy landscape of vertex models is non-analytic, or ``cuspy". Unlike most particulate matter, in vertex models there is a decoupling between the linear response and the non-linear response. In this specific case, the energy landscape for the ordered tissue is perfectly flat in a ball of radius $l_0-l^*$ from the ordered ground state, and then rises sharply from zero starting at $l^*$.  This cuspy landscape has already been identified and implicated in other processes in 2D vertex models, including unexpectedly sharp interfaces between two tissue types~\cite{Sussman2018}.  In that work, it was demonstrated that the cuspy landscape is independent of the exact form of the model (i.e. Vertex vs. Voronoi). It was also argued that we should expect non-analytic behavior in any model with topological interactions between cells, where neighbors are defined as those that share an edge, instead of metric interactions, where neighbors are defined by how far apart they are. Additional work by some of us confirms that many types of models with topological connections, including underconstrained fiber networks, exhibit universal behavior governed by an underlying geometric incompatibility~\cite{Merkel2019}. Therefore, it is interesting to conjecture that any model with topological interactions, such as those for bird flocks and certain biomimetic- and meta- materials, might have similar features with deep connections between ordered and disordered states. 

Another hint at this deep connection comes from beautiful work by Moshe \textit{et al}~\cite{Moshe2018}, who develop an analytic model based on intrinsic metrics for periodic vertex lattices.  In that work, they focus on an elastic model with no rearrangements where deformations from target metrics are quadratically penalized, and they predict from first principles that for $s_0 > 3.72$, the energy landscape in the space of metrics is also perfectly flat. It would be interesting to see if extensions of that framework might be able to account for nonlinearities, and perhaps find some non-analyticity in the space of metrics, in order to explain non-linear stabilization in real space.  If possible, our work suggests that may be a productive path towards a first-principles prediction of rigidity in a disordered system, which would be very exciting.

A related manuscript that also highlights the importance of flat energy landscapes in ordered and disordered cellular systems is the work by Noll \textit{et al}~\cite{Noll2017} on isogonal modes in force-balanced tension networks.  In that work, a different version of the vertex model, without a $P^2$ term in the energy functional to act as a restoring force, is coupled with myosin dynamics. The form of feedback chosen to model the myosin dynamics, which has recently been confirmed in experiments on fruit flies~\cite{Streichan2018}, introduces a different type of restoring force that permits mechanically stable cellular networks. Although their myosin-feedback model and our standard ordered vertex model both possess zero-energy linear modes, their zero modes must be angle-preserving while perturbations associated with our T1 transitions explicitly change angles.  Given this, it would be interesting to study how the functional form of restoring forces in the energy functional for vertex models impacts the linear and nonlinear stability of cellular networks.

Finally, this work focuses on vertex models in the absence of fluctuations, i.e. at zero temperature. An interesting future direction would be to study how the effective linear response and nonlinear stability changes as a function of temperature or self-propulsion.  For example, in ordered systems with $3.72 < s_0 < 3.81$ one might expect that at low temperatures, fluctuations typically remain small and only probe the linear regime with no shear modulus. At higher temperatures fluctuations would regularly probe the nonlinear response, so the effective linear response has a finite shear modulus. Moreover, active or driven fluctuations with a persistence time would sample these non-linear regions in different ways, perhaps leading to very rich behavior.  

Given the existence and importance of ordered cellular networks in epithelial layers in developmental systems ranging from fruit flies to vertebrates, our results might impact how we think about their form and function. Specifically, we suggest that the mechanical properties of such tissues are quite exotic, with interesting nonlinearities and possible fluctuation-induced solidification. We speculate that perhaps some biological tissues tune themselves to take advantage of these interesting properties and functions.

\subsubsection*{Acknowledgement}
 We thank Jennifer Schwarz for fruitful discussions. This work was primarily supported by NSF-POLS-1607416 and NSF DMR-1460784 (REU). MLM and PS acknowledge additional support from Simons Grant No. 454947 and NSF-DMR -1352184, and MLM and GET acknowledge support from Simons Grant No. 446222 and and NIH R01GM117598.

%\subsection{\label{sec:citeref}Citations and References}
%\bibliographystyle{plain}
\bibliographystyle{unsrtnat}
\bibliography{main.bib}

\begin{thebibliography}{34}
\providecommand{\natexlab}[1]{#1}
\providecommand{\url}[1]{\texttt{#1}}
\expandafter\ifx\csname urlstyle\endcsname\relax
  \providecommand{\doi}[1]{doi: #1}\else
  \providecommand{\doi}{doi: \begingroup \urlstyle{rm}\Url}\fi

\bibitem[Farhadifar et~al.(2007)Farhadifar, R{\"{o}}per, Aigouy, Eaton, and
  J{\"{u}}licher]{Farhadifar2007a}
Reza Farhadifar, Jens~Christian R{\"{o}}per, Benoit Aigouy, Suzanne Eaton, and
  Frank J{\"{u}}licher.
\newblock {The Influence of Cell Mechanics, Cell-Cell Interactions, and
  Proliferation on Epithelial Packing}.
\newblock \emph{Curr. Biol.}, 17\penalty0 (24):\penalty0 2095--2104, 2007.
\newblock ISSN 09609822.
\newblock \doi{10.1016/j.cub.2007.11.049}.

\bibitem[Staple et~al.(2010)Staple, Farhadifar, R{\"{o}}per, Aigouy, Eaton, and
  J{\"{u}}licher]{Staple2010}
D.~B. Staple, R.~Farhadifar, J.~C. R{\"{o}}per, B.~Aigouy, S.~Eaton, and
  F.~J{\"{u}}licher.
\newblock {Mechanics and remodelling of cell packings in epithelia}.
\newblock \emph{Eur. Phys. J. E}, 33\penalty0 (2):\penalty0 117--127, 2010.
\newblock ISSN 12928941.
\newblock \doi{10.1140/epje/i2010-10677-0}.

\bibitem[Zhang et~al.(2017)Zhang, Xu, Lee, Zhu, Wu, Liao, Zhang, Sun, Mogilner,
  Losert, Pan, Lin, Xu, and Zhao]{Zhang2017}
Yan Zhang, Guoqing Xu, Rachel~M. Lee, Zijie Zhu, Jiandong Wu, Simon Liao, Gong
  Zhang, Yaohui Sun, Alex Mogilner, Wolfgang Losert, Tingrui Pan, Francis Lin,
  Zhengping Xu, and Min Zhao.
\newblock {Collective cell migration has distinct directionality and speed
  dynamics}.
\newblock \emph{Cell. Mol. Life Sci.}, 2017.
\newblock ISSN 14209071.
\newblock \doi{10.1007/s00018-017-2553-6}.

\bibitem[Serra-Picamal et~al.(2012)Serra-Picamal, Conte, Vincent, Anon, Tambe,
  Bazellieres, Butler, Fredberg, and Trepat]{Serra-Picamal2012}
Xavier Serra-Picamal, Vito Conte, Romaric Vincent, Ester Anon, Dhananjay~T.
  Tambe, Elsa Bazellieres, James~P. Butler, Jeffrey~J. Fredberg, and Xavier
  Trepat.
\newblock {Mechanical waves during tissue expansion}.
\newblock \emph{Nat. Phys.}, 2012.
\newblock ISSN 17452481.
\newblock \doi{10.1038/nphys2355}.

\bibitem[Mongera et~al.(2018)Mongera, Rowghanian, Gustafson, Shelton,
  Kealhofer, Carn, Serwane, Lucio, Giammona, and Camp{\`{a}}s]{Mongera2018}
Alessandro Mongera, Payam Rowghanian, Hannah~J. Gustafson, Elijah Shelton,
  David~A. Kealhofer, Emmet~K. Carn, Friedhelm Serwane, Adam~A. Lucio, James
  Giammona, and Otger Camp{\`{a}}s.
\newblock {A fluid-to-solid jamming transition underlies vertebrate body axis
  elongation}, 2018.
\newblock ISSN 14764687.

\bibitem[Park et~al.(2015)Park, Kim, Bi, Mitchel, Qazvini, Tantisira, Park,
  McGill, Kim, Gweon, Notbohm, Steward, Burger, Randell, Kho, Tambe, Hardin,
  Shore, Israel, Weitz, Tschumperlin, Henske, Weiss, Manning, Butler, Drazen,
  and Fredberg]{Park2015b}
Jin~Ah Park, Jae~Hun Kim, Dapeng Bi, Jennifer~A. Mitchel, Nader~Taheri Qazvini,
  Kelan Tantisira, Chan~Young Park, Maureen McGill, Sae~Hoon Kim, Bomi Gweon,
  Jacob Notbohm, Robert Steward, Stephanie Burger, Scott~H. Randell, Alvin~T.
  Kho, Dhananjay~T. Tambe, Corey Hardin, Stephanie~A. Shore, Elliot Israel,
  David~A. Weitz, Daniel~J. Tschumperlin, Elizabeth~P. Henske, Scott~T. Weiss,
  M.~Lisa Manning, James~P. Butler, Jeffrey~M. Drazen, and Jeffrey~J. Fredberg.
\newblock {Unjamming and cell shape in the asthmatic airway epithelium}.
\newblock \emph{Nat. Mater.}, 2015.
\newblock ISSN 14764660.
\newblock \doi{10.1038/nmat4357}.

\bibitem[Yamada et~al.(2005)Yamada, Pokutta, Drees, Weis, and
  Nelson]{Yamada2005}
Soichiro Yamada, Sabine Pokutta, Frauke Drees, William~I. Weis, and W.~James
  Nelson.
\newblock {Deconstructing the cadherin-catenin-actin complex}.
\newblock \emph{Cell}, 2005.
\newblock ISSN 00928674.
\newblock \doi{10.1016/j.cell.2005.09.020}.

\bibitem[Ma{\^{i}}tre et~al.(2012)Ma{\^{i}}tre, Berthoumieux, Krens, Salbreux,
  J{\"{u}}licher, Paluch, and Heisenberg]{Maitre2012a}
Jean~L{\'{e}}on Ma{\^{i}}tre, H{\'{e}}l{\`{e}}ne Berthoumieux, Simon
  Frederik~Gabriel Krens, Guillaume Salbreux, Frank J{\"{u}}licher, Ewa Paluch,
  and Carl~Philipp Heisenberg.
\newblock {Adhesion functions in cell sorting by mechanically coupling the
  cortices of adhering cells}.
\newblock \emph{Science (80-. ).}, 2012.
\newblock ISSN 10959203.
\newblock \doi{10.1126/science.1225399}.

\bibitem[Nagai and Honda(2001)]{Nagai2001a}
Tatsuzo Nagai and Hisao Honda.
\newblock {A dynamic cell model for the formation of epithelial tissues}.
\newblock \emph{Philos. Mag. B Phys. Condens. Matter; Stat. Mech. Electron.
  Opt. Magn. Prop.}, 2001.
\newblock ISSN 13642812.
\newblock \doi{10.1080/13642810108205772}.

\bibitem[Chiou et~al.(2012)Chiou, Hufnagel, and Shraiman]{Chiou2012}
Kevin~K. Chiou, Lars Hufnagel, and Boris~I. Shraiman.
\newblock {Mechanical stress inference for two dimensional cell arrays}.
\newblock \emph{PLoS Comput. Biol.}, 2012.
\newblock ISSN 1553734X.
\newblock \doi{10.1371/journal.pcbi.1002512}.

\bibitem[Fletcher et~al.(2017)Fletcher, Cooper, and Baker]{Fletcher2017}
Alexander~G. Fletcher, Fergus Cooper, and Ruth~E. Baker.
\newblock {Mechanocellular models of epithelial morphogenesis}, 2017.
\newblock ISSN 14712970.

\bibitem[Merkel and Manning(2017)]{Merkel2017}
Matthias Merkel and M.~Lisa Manning.
\newblock {Using cell deformation and motion to predict forces and collective
  behavior in morphogenesis}, 2017.
\newblock ISSN 10963634.

\bibitem[Manning et~al.(2010)Manning, Foty, Steinberg, and
  Schoetz]{Manning2010b}
M.~Lisa Manning, Ramsey~A. Foty, Malcolm~S. Steinberg, and Eva-Maria Schoetz.
\newblock {Coaction of intercellular adhesion and cortical tension specifies
  tissue surface tension}.
\newblock \emph{Proc. Natl. Acad. Sci.}, 2010.
\newblock ISSN 0027-8424.
\newblock \doi{10.1073/pnas.1003743107}.

\bibitem[Nnetu et~al.(2012)Nnetu, Knorr, K{\"{a}}s, and Zink]{Nnetu2012}
Kenechukwu~David Nnetu, Melanie Knorr, Josef K{\"{a}}s, and Mareike Zink.
\newblock {The impact of jamming on boundaries of collectively moving
  weak-interacting cells}.
\newblock \emph{New J. Phys.}, 2012.
\newblock ISSN 13672630.
\newblock \doi{10.1088/1367-2630/14/11/115012}.

\bibitem[Bi et~al.(2015)Bi, Lopez, Schwarz, and Manning]{Bi2015c}
Dapeng Bi, J.~H. Lopez, J.~M. Schwarz, and M.~Lisa Manning.
\newblock {A density-independent rigidity transition in biological tissues}.
\newblock \emph{Nat. Phys.}, 11\penalty0 (12):\penalty0 1074--1079, 2015.
\newblock ISSN 17452481.
\newblock \doi{10.1038/nphys3471}.

\bibitem[Classen et~al.(2005)Classen, Anderson, Marois, and
  Eaton]{Classen2005a}
Anne~Kathrin Classen, Kurt~I. Anderson, Eric Marois, and Suzanne Eaton.
\newblock {Hexagonal packing of Drosophila wing epithelial cells by the planar
  cell polarity pathway}.
\newblock \emph{Dev. Cell}, 2005.
\newblock ISSN 15345807.
\newblock \doi{10.1016/j.devcel.2005.10.016}.

\bibitem[McKenzie et~al.(2004)McKenzie, Krupin, and Kelley]{McKenzie2004}
Erynn McKenzie, Alison Krupin, and Matthew~W. Kelley.
\newblock {Cellular Growth and Rearrangement during the Development of the
  Mammalian Organ of Corti}.
\newblock \emph{Dev. Dyn.}, 2004.
\newblock ISSN 10588388.
\newblock \doi{10.1002/dvdy.10500}.

\bibitem[Tardieu(2002)]{Tardieu2002}
A~Tardieu.
\newblock {Eye Lens Proteins And Transparency: From Light Transmission Theory
  To Solution X-Ray Structural Analysis}.
\newblock \emph{Annu. Rev. Biophys. Biomol. Struct.}, 2002.
\newblock ISSN 10568700.
\newblock \doi{10.1146/annurev.biophys.17.1.47}.

\bibitem[Cooper et~al.(2008)Cooper, Son, Komlos, Sun, Kleiman, and
  Zhou]{Cooper2008}
Margaret~A Cooper, Alexander~I Son, Daniel Komlos, Yuhai Sun, Norman~J Kleiman,
  and Renping Zhou.
\newblock {Loss of ephrin-A5 function disrupts lens fiber cell packing and
  leads to cataract}.
\newblock \emph{Proc. Natl. Acad. Sci.}, 2008.
\newblock \doi{10.1073/pnas.0808987105}.

\bibitem[Pontani et~al.(2016)Pontani, Jorjadze, and Brujic]{Pontani2016}
Lea~Laetitia Pontani, Ivane Jorjadze, and Jasna Brujic.
\newblock {Cis and Trans Cooperativity of E-Cadherin Mediates Adhesion in
  Biomimetic Lipid Droplets}.
\newblock \emph{Biophys. J.}, 2016.
\newblock ISSN 15420086.
\newblock \doi{10.1016/j.bpj.2015.11.3514}.

\bibitem[Sussman et~al.(2018)Sussman, Paoluzzi, {Cristina Marchetti}, and {Lisa
  Manning}]{Sussman2018}
Daniel~M. Sussman, M.~Paoluzzi, M.~{Cristina Marchetti}, and M.~{Lisa Manning}.
\newblock {Anomalous glassy dynamics in simple models of dense biological
  tissue}.
\newblock \emph{Epl}, 121\penalty0 (3), 2018.
\newblock ISSN 12864854.
\newblock \doi{10.1209/0295-5075/121/36001}.

\bibitem[Kamien and Liu(2007)]{Kamien2007}
Randall~D. Kamien and Andrea~J. Liu.
\newblock {Why is random close packing reproducible?}
\newblock \emph{Phys. Rev. Lett.}, 2007.
\newblock ISSN 00319007.
\newblock \doi{10.1103/PhysRevLett.99.155501}.

\bibitem[Teleman et~al.(2007)Teleman, Hufnagel, Rouault, Shraiman, and
  Cohen]{Teleman2007}
A.~A. Teleman, L.~Hufnagel, H.~Rouault, B.~I. Shraiman, and S.~M. Cohen.
\newblock {On the mechanism of wing size determination in fly development}.
\newblock \emph{Proc. Natl. Acad. Sci.}, 2007.
\newblock ISSN 0027-8424.
\newblock \doi{10.1073/pnas.0607134104}.

\bibitem[Hilgenfeldt et~al.(2008)Hilgenfeldt, Erisken, and
  Carthew]{Hilgenfeldt2008}
S.~Hilgenfeldt, S.~Erisken, and R.~W. Carthew.
\newblock {Physical modeling of cell geometric order in an epithelial tissue}.
\newblock \emph{Proc. Natl. Acad. Sci.}, 2008.
\newblock ISSN 0027-8424.
\newblock \doi{10.1073/pnas.0711077105}.

\bibitem[Fletcher et~al.(2014)Fletcher, Osterfield, Baker, and
  Shvartsman]{Fletcher2014}
Alexander~G. Fletcher, Miriam Osterfield, Ruth~E. Baker, and Stanislav~Y.
  Shvartsman.
\newblock {Vertex models of epithelial morphogenesis}, 2014.
\newblock ISSN 15420086.

\bibitem[Bi et~al.(2014)Bi, Lopez, Schwarz, and {Lisa Manning}]{Bi2014}
Dapeng Bi, Jorge~H. Lopez, J.~M. Schwarz, and M.~{Lisa Manning}.
\newblock {Energy barriers and cell migration in densely packed tissues}.
\newblock \emph{Soft Matter}, 2014.
\newblock ISSN 1744683X.
\newblock \doi{10.1039/c3sm52893f}.

\bibitem[Sussman(2017)]{Sussman2017}
Daniel~M. Sussman.
\newblock {cellGPU: Massively parallel simulations of dynamic vertex models}.
\newblock \emph{Comput. Phys. Commun.}, 219:\penalty0 400--406, oct 2017.
\newblock ISSN 0010-4655.
\newblock \doi{10.1016/J.CPC.2017.06.001}.
\newblock URL
  \url{https://www.sciencedirect.com/science/article/pii/S0010465517301832}.

\bibitem[Bitzek et~al.(2006)Bitzek, Koskinen, G{\"{a}}hler, Moseler, and
  Gumbsch]{Bitzek2006}
Erik Bitzek, Pekka Koskinen, Franz G{\"{a}}hler, Michael Moseler, and Peter
  Gumbsch.
\newblock {Structural relaxation made simple}.
\newblock \emph{Phys. Rev. Lett.}, 2006.
\newblock ISSN 00319007.
\newblock \doi{10.1103/PhysRevLett.97.170201}.

\bibitem[Sahu et~al.(2019)Sahu, Sussman, Marchetti, Manning, and
  Schwarz]{Sahu2019}
Preeti Sahu, Daniel~M. Sussman, M.~Cristina Marchetti, M.~Lisa Manning, and
  J.~M. Schwarz.
\newblock {Large-scale mixing and small-scale demixing in a confluent model for
  biological tissues}.
\newblock 2019.
\newblock URL \url{https://arxiv.org/pdf/1905.00657v1.pdf}.

\bibitem[Kim et~al.(2018)Kim, Wang, and Hilgenfeldt]{Kim2018}
Sangwoo Kim, Yiliang Wang, and Sascha Hilgenfeldt.
\newblock {Universal Features of Metastable State Energies in Cellular Matter}.
\newblock \emph{Phys. Rev. Lett.}, 2018.
\newblock ISSN 10797114.
\newblock \doi{10.1103/PhysRevLett.120.248001}.

\bibitem[Merkel et~al.(2019)Merkel, Baumgarten, Tighe, and Manning]{Merkel2019}
Matthias Merkel, Karsten Baumgarten, Brian~P. Tighe, and M.~Lisa Manning.
\newblock {A minimal-length approach unifies rigidity in underconstrained
  materials}.
\newblock \emph{Proc. Natl. Acad. Sci.}, 2019.
\newblock ISSN 0027-8424.
\newblock \doi{10.1073/pnas.1815436116}.

\bibitem[Moshe et~al.(2018)Moshe, Bowick, and Marchetti]{Moshe2018}
Michael Moshe, Mark~J. Bowick, and M.~Cristina Marchetti.
\newblock {Geometric Frustration and Solid-Solid Transitions in Model 2D
  Tissue}.
\newblock \emph{Phys. Rev. Lett.}, 120\penalty0 (26):\penalty0 268105, 2018.
\newblock ISSN 10797114.
\newblock \doi{10.1103/PhysRevLett.120.268105}.
\newblock URL \url{https://doi.org/10.1103/PhysRevLett.120.268105}.

\bibitem[Noll et~al.(2017)Noll, Mani, Heemskerk, Streichan, and
  Shraiman]{Noll2017}
Nicholas Noll, Madhav Mani, Idse Heemskerk, Sebastian~J. Streichan, and
  Boris~I. Shraiman.
\newblock {Active tension network model suggests an exotic mechanical state
  realized in epithelial tissues}.
\newblock \emph{Nat. Phys.}, 2017.
\newblock ISSN 17452481.
\newblock \doi{10.1038/nphys4219}.

\bibitem[Streichan et~al.(2018)Streichan, Lefebvre, Noll, Wieschaus, and
  Shraiman]{Streichan2018}
Sebastian~J Streichan, Matthew~F Lefebvre, Nicholas Noll, Eric~F Wieschaus, and
  Boris~I Shraiman.
\newblock {Global morphogenetic flow is accurately predicted by the spatial
  distribution of myosin motors}.
\newblock \emph{Elife}, 2018.
\newblock \doi{10.7554/elife.27454}.

\end{thebibliography}

\beginsupplement

\section{\label{sec:SI} Electronic Supplementary Information}

\subsection*{Comparison to disordered packings}

\begin{figure}[!htbp]
\centering
\includegraphics[width=1.05\columnwidth]{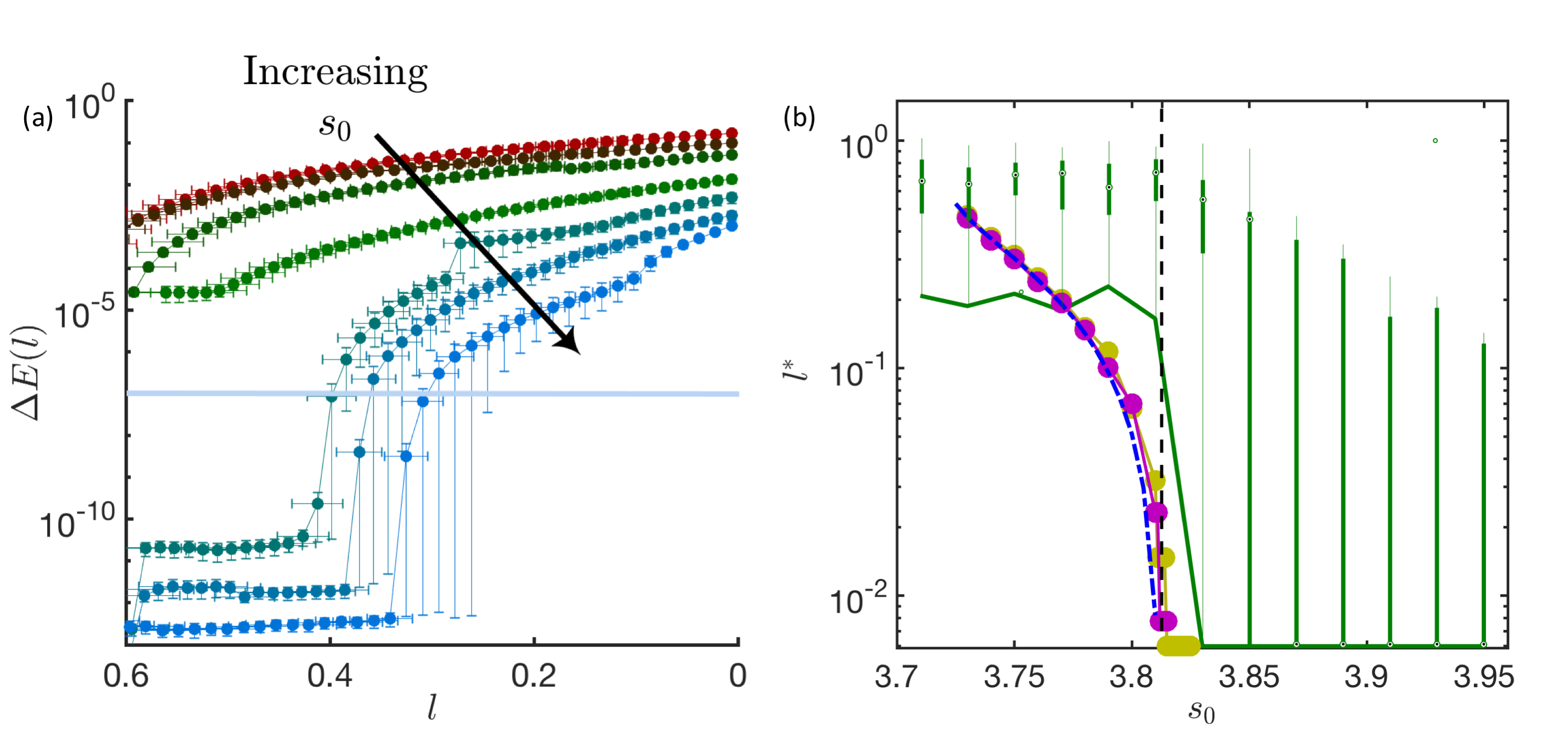}
\caption{\textbf{ Many-cell disordered energy profile}: (a) In a disordered system of 90 cells, a randomly chosen edge undergoes a T1 transition for 50 different initializations. In this process, the relative energy of the tissue, $\Delta E (l)$, is plotted against the shrinking T1 edgelength $l$ for increasing values of $s_0$ (3.71 to 3.95 in steps of 0.04) varying from red (3.71) to green (3.83) to blue (3.95). The cut-off for the energy is shown by a horizontal pale blue line for reference. (b) Critical edgelength $l^*$ plotted against $s_0$ is superimposed for both- many-cell (yellow circles) and 4-cell systems (magenta circles). The analytical prediction from the geometric mechanism explained in the text is shown in blue dashed line. The dark green box and whisker plot in blue shows the $l^*$ distribution in disordered systems.}
\label{fig:disordered}
\end{figure}

To compare our results on ordered systems to those in disordered systems, we investigate the onset of non-linearities in disordered systems. Specifically we study the properties of systems with shrinking T1 edges from 50 different initializations. Unlike ordered simulations, these systems do not start from their ground state. They are equilibrated from a uniform random initial configuration and during the initial equilibration process we use a higher $l_c=0.15$ which allows the system to explore more states on the trajectory towards a local energy minimum. Once the system has arrived at a mechanically stable state, we start the same process of shrinking a random edge to a length as small as $l_c=0.006$. Since this initial energy now is not necessarily zero for $s_0>3.722$, we look at the relative energy $\Delta E(l)$ from initial state at every edgelength. We bin every T1 edgelength into 40 bins. To look at the average trend of these profiles as a function of increasing shape, we average $\Delta E(l)$ for every bin. We have used the same color scheme as in previous plots, and so one can see that for the disordered case, the energy remains high at all values of $l$ throughout the entire range explored previously ($s_0$ in 3.71-3.83). For $s_0>3.83$, the average energy drops precipitously at an $l$ value smaller than the average. It is important to note that we have focused on average values in Fig.~\ref{fig:disordered}(a), but there are large fluctuations in edgelength due to the disorder, and the system will be unstable if any edge in the system can move at zero cost. Therefore, to find the $l^*$ for a given configuration, we should focus on the lowest $l^*$, not the average, as shown in Fig.~\ref{fig:disordered}(b).

For this energy profile, we use the same energy cut-off and identify the critical edgelength $l^*$ for an edge in every  ensemble. We find that in general this ensemble exhibits a wide distribution of $l^*$s because disordered systems have a variety of edgelengths. Therefore, we represent this data using a box and whisker plot as shown in Fig.~\ref{fig:disordered}(b). 

As previous work suggests that linear curvature does not vanish until approximately 3.81 for disordered systems, for $s_0 < 3.81$ one should expect the energy to grow as soon as the edge starts shrinking, so that $l^*=l_0$. As for $s_0 > 3.81$ the system is fluid so it should be possible for some edges to shrink to zero length at no energy cost, so that $l^*=0$.

As shown in Fig\ref{fig:disordered}, our data is in line with these expectation. For $s_0 < 3.81$,  $l^*$ is large and approximately equal to $l_0$, while for $s_0 > 3.81$, there are some edges for which $l^*$ approaches zero, resulting in a near discontinuity in the plot.

\bigskip
\bigskip
\bigskip
\bigskip
\bigskip
\bigskip
\bigskip
\bigskip
\bigskip
\bigskip
\bigskip
\bigskip
\bigskip
\bigskip

\end{document}